\def\year{2024}\relax
\documentclass[letterpaper]{article} 
\usepackage{aaai22}  

\usepackage{times}  
\usepackage{helvet}  
\usepackage{courier}  
\usepackage[hyphens]{url}  
\usepackage{graphicx} 
\usepackage{natbib}  
\usepackage{caption} 
\usepackage{multirow, makecell}
\DeclareCaptionStyle{ruled}{labelfont=normalfont,labelsep=colon,strut=off} 
\usepackage{siunitx}
\newcolumntype{L}[1]{>{\raggedright\let\newline\\\arraybackslash\hspace{0pt}}m{#1}}
\newcolumntype{C}[1]{>{\centering\let\newline\\\arraybackslash\hspace{0pt}}m{#1}}
\newcolumntype{R}[1]{>{\raggedleft\let\newline\\\arraybackslash\hspace{0pt}}m{#1}}

\usepackage{algorithm}
\usepackage{algorithmic}
\usepackage{booktabs}
\usepackage{newfloat}
\usepackage{listings}
\usepackage{siunitx}
\newcolumntype{P}[1]{>{\centering\arraybackslash}p{#1}}

\usepackage{xcolor}

\usepackage[hang,flushmargin]{footmisc}
\floatstyle{ruled}
\newfloat{listing}{tb}{lst}{}
\floatname{listing}{Listing}
\title{Facebook Political Ads and Accountability:\\Outside Groups Are Most Negative, Especially When Hiding Donors}
\author {
    Shomik Jain\textsuperscript{\rm 1} and
    Abby K. Wood\textsuperscript{\rm 2}
}
\affiliations {
    \textsuperscript{\rm 1} Institute for Data, Systems, and Society, Massachusetts Institute of Technology\\
    \textsuperscript{\rm 2} Gould School of Law, University of Southern California\\
    shomikj@mit.edu, awood@law.usc.edu
}

\begin{document}

\maketitle

\begin{abstract}
The emergence of online political advertising has come with little regulation, allowing political advertisers on social media to avoid accountability. We analyze how transparency and accountability deficits caused by dark money and disappearing groups relate to the sentiment of political ads on Facebook. We obtained 430,044 ads with FEC-registered advertisers from Facebook’s ad library that ran between August-November 2018. We compare ads run by candidates, parties, and outside groups, which we classify by (1) their donor transparency (dark money or disclosed) and (2) the group's permanence (only FEC-registered in 2018 or persistent across cycles). The most negative advertising came from dark money and disappearing outside groups, which were mostly corporations or 501(c) organizations. However, only dark money was associated with a significant decrease in ad sentiment. These results suggest that accountability for political speech matters for advertising tone, especially in the context of affective polarization on social media.
\end{abstract}

\section{Introduction}

Online political advertisements play an increasing and controversial role in U.S. elections. Facebook generated over \$100 billion from advertising revenue in 2022. However, the platform has come under scrutiny for enabling advertising that spreads disinformation and furthers societal polarization~\citep{fowler_apsr, kriess_mcgregor_microtargeting}. For example, Russia targeted undecided voters in the 2016 election with negative and deceptive ads on Facebook to sow discord and help elect President Trump~\citep{ribeiro_microtargeting, eady2023russia}. These recent events have fueled discussions about a broader issue: the limited transparency surrounding online political ads~\citep{edelson_transparency, kim_darkMoney, wood_sclr}. In this work, we study how campaign finance disclosure and group permanence, two avenues for accountability, relate to the sentiment of Facebook advertising. 

We focus on sentiment because affective polarization -- the tendency to view opposing partisans negatively -- has increased alongside negative political speech online. Negative advertising and criticism of the opposing party on social media has been shown to exacerbate partisans' negative perceptions of their opponents~\citep{iyengar_2019, lau2017effect, suhay_2018, tucker_review}. Ahead of the 2022 election, 72\% of Republicans and 63\% of Democrats said people in the opposing party are more immoral than other Americans, up from less than half in each party in 2016~\citep{pew_affective}. This increasing polarization poses societal risks, such as fostering hate speech, changing public support for policies, and threatening democratic norms~\citep{druckman2021affective, hate_speech, pnas_polarization}. Moreover, there may also be a link between negativity and disinformation. Early research on the relationship suggests that negativity predicts so-called ``fake news''~\citep{Bozarth_Budak_2020, jiang_misinf_sent, vicario_misinf_sent}. On the other hand, negative political ads \textit{can} be informative~\citep{dowling_negative}, but their persuasion effects are small across many contexts~\citep{science_attackAdsEffects}. The persuasion effects of ``going negative'' may be conditional on several factors, such as the message content and target~\citep{fridkin_2011, blackwell_2013}. Given the growing prevalence of political advertising on social media, as well as increasing affective polarization and disinformation, it is important to understand how the sentiment of online advertising relates to variations in advertiser transparency. 

Current federal campaign finance laws make it difficult to trace the money behind many political advertisements, especially those on social media platforms such as Facebook~\citep{wood_arlss}. In this era of unlimited outside spending ushered in by \textit{Citizens United v. FEC} and \textit{SpeechNow v. FEC}, outside groups that make political expenditures independently of candidate committees have steadily increased spending in each election cycle. \textit{Citizens United} lifted a ban on these so-called independent expenditures from corporations' general treasuries, and the Federal Election Commission (FEC) has largely failed to require any donor disclosure for them. The result is dark money: political spending from anonymous sources. Dark money can also fund ``Super PACs,'' or political action committees that only make independent expenditures. Super PACs are  required to disclose their donors. However, after the \textit{SpeechNow} court ruled that Super PACs have no contribution limits, they began to receive unlimited contributions from groups without disclosure requirements, effectively making Super PAC donors untraceable in many cases. The 2020 election set records with over \$3 billion in outside spending and at least \$1 billion from dark money groups~\citep{crp_spending}. In a different set of accountability limitations, some outside groups can also disappear after running ads in only one election cycle. While any outside group can disappear, limited liability corporations (LLCs) are often set up for short-term spending because they face little regulation. These disappearing outside groups are less accountable to the public than candidates and parties, who are recognizable and cannot disappear~\citep{herrnson_superPAC, stan_popup}. When advertisers hide their donors, disappear after elections, or both, it is impossible for the public to hold them accountable for their political speech.

How do dark money and group persistence relate to the tone of political advertisers on Facebook? The existing literature from television and radio ads suggests that political advertisers with less transparency -- which includes many kinds of outside groups -- may be more willing to address divisive issues and run negative messages~\citep{chand_negative, dowling_negative, fowler_negative, stan_popup}. The ``backlash'' effect, where voters tend to have lower support for the sponsors of negative ads, helps explain the phenomenon of campaigns and parties ``outsourcing'' negativity to outside groups~\citep{dowling_negative}. However, voter backlash against negative ads run by unfamiliar advertisers, like new outside groups or those with hidden donors, is smaller than the backlash effect for more familiar advertisers such as candidates and parties~\citep{dowling_wichowsky_cf_disclosure, dowling_wichowsky_negative}. Of course, this makes avoiding transparency attractive to advertisers. It also explains why why outside groups may prefer to reconstitute under a different name between election cycles, to avoid becoming familiar to voters. Theories about how advertiser accountability relates to ad sentiment have not been tested in the online context, though in our discussion of related literature below, we review several works that have analyzed the sentiment and content of online political ads more generally. 

This work analyzes the sentiment of Facebook political ads from the 2018 U.S. election. We compare candidates, parties, and outside groups based on their donor disclosure and persistence across election cycles. We hypothesize that insulation from accountability breeds advertising negativity. Because accountability in the political arena comes through both donor disclosure and advertiser permanence, we expect that ads from organizations lacking either of these features will be more negative. Our results partially support our hypothesis: We find that the most negative advertising comes from outside groups that are funded by dark money, particularly if they are disappearing and only registered with the FEC for the 2018 cycle. However, as we discuss below, the main association between ad sentiment and advertiser accountability is through donor disclosure. Dark money ads are significantly more negative than those with disclosed donors, regardless of group permanence.

\section{Related Literature}

Many prior works have studied online political ads, with analyses spanning advertising content, influence, and transparency~\citep{fowler2020book, pierri2023political, capozzi2023thin, aisenpreis2023us, coelho2023propaganda}. Several of these works focus on Facebook. Facebook is the top platform for political advertising by both the amount of money spent and the number of voters reached~\citep{edelson_transparency}. The release of the Facebook ad library has also enabled more granular research about specific ads and advertiser types, despite its many limitations~\citep{edelson_api}. In this section, we review related studies of Facebook political ads that analyze (1) advertiser type, (2) advertiser transparency, (3) ad targeting and (4) ad sentiment. 

Outside groups make up a much higher share of political advertisers on Facebook when compared to television and radio advertising. \citet{edelson_transparency} conducted one of the first studies of Facebook's library, and found that candidates and parties only accounted for 17\% and 8\% of advertising spend, respectively. They also found several instances of advertising by quasi for-profit media companies (often LLCs) that appeared to exist for the sole purpose of spreading political messaging. Advertiser type and advertiser transparency are closely related. For example in federal elections, if an advertiser is not registered with the FEC, then its donors will not be disclosed via that mechanism. A recent audit found over 100,000 ads with political content that are not included in Facebook's library, almost entirely from advertisers not registered with the Federal Election Commission (FEC)~\citep{le2022audit}. \citet{kim_darkMoney} found that these non-FEC registered advertisers used more divisive issue campaigns, and ran 4 times the number of ads as FEC-registered advertisers.

Scholars are also making progress on studying ad targeting, which represents another mechanism for avoiding accountability. \citet{calvo2021global} examined data from Spanish elections in 2019, finding that parties implemented targeting strategically to their bases both in terms of outreach and content. In a related study, \citet{capozzi2021clandestino} found that anti-immigration ads in Italy were more likely to be targeted towards men. However, there is mixed support for the relationship between the ``toxicity'' of an ad and the decision to narrow its audience via targeting, though this does vary by the type of advertiser (whether it is the campaign itself or an outside group)~\citep{votta2023going}. Furthermore, \citet{ali2021ad} reveal how online platforms can further optimize ad delivery beyond the intended targets of an advertiser. They also find that platforms financially disincentivize advertisers from targeting audiences that are not ideologically-aligned. These results show how online platforms may contribute to polarization by limiting the exposure of users to contrasting viewpoints.

In addition to these accountability deficits, several previous works have also analyzed the sentiment of Facebook political ads, although not in relation to advertiser transparency (as is the focus of this work). \citet{fowler_apsr, fowler2022midterms} showed that candidates are more positive in Facebook ads when compared to those they run on television. In particular, they found that candidates engage in less attacking of their opponents and more promotion of themselves on Facebook. The sentiment of candidate ads was also found to vary based on their political affiliation and the ad's subject. For example, ads about climate change were more negative across Republican candidates than Democratic ones~\citet{aisenpreis2023us}. Moreover, Russian propagandists leveraged affective triggers differently across the 2016 election cycle: ads were more negative prior to the election and more positive immediately afterwards~\citep{alvarez2020good}. Across these ads run by foreign actors, the ones with the most impressions were more likely to be negative~\citep{dutt2019senator}. In European elections, \citet{kruschinski2022posting} discovered that opposition parties are more likely to go negative than incumbents. The majority of Facebook ads in several European elections were also positive across all advertiser types~\citep{lopez2021microtargeted}.

Despite a strong and growing body of research about online political advertising, an open question remains as to how deficits in accountability -- specifically dark money and advertiser disappearance -- relate to the sentiment of online political ads. This work aims to address this gap in the literature.

\section{Data and Methods}

\subsection{Data Sources}

We collected the complete set of ads in Facebook's political ad library for the 3-month window before the 2018 U.S. midterm election. We acquired this data through a license with \citeauthor{harmony_labs}, a third-party organization that hosts a database of scraped data from \citeauthor{facebook_adLibrary}'s ad library. Several issues with Facebook's API, such as extreme limitations on query rates and formats, prevented us from obtaining the data directly as well as for other election years~\citep{edelson_api, edelson_nyt}. To categorize advertiser types and accountability, we matched Facebook advertiser names and political advertiser data from the \citeauthor{fec_committees}, \citeauthor{irs_501c}, \citeauthor{crp_tracing}, and \citeauthor{opencorporates}. We determined Facebook advertiser names using the ``Paid for by'' label associated with each ad, and manually verified matches with the political advertiser data. This procedure allowed us to classify advertisers as candidates, parties, or outside groups. We further grouped advertisers as Democratic (or having a liberal viewpoint) or Republican (or having a conservative viewpoint) using their FEC-registration (for candidates and parties) or by manually labeling their party-affiliation\footnote{We did not assign a party-affiliation to some advertisers ($<$1\% of ads) if they only ran non-political ads or supported both parties.} based on the content of their ads (for outside groups).

\subsection{Federal Scope} We limit the scope of our analysis to 430,044 ads from the 1,277 advertisers that (1) registered with the FEC, and (2) for outside groups, had their amount of donor disclosure recorded by the CRP. We focus on FEC-registered advertisers because they represent the realm of advertisers subject to federal regulation~\citep{wood_arlss}. Most political advertisers who spend above a regulatory minimum in federal elections must register with the FEC, even if they do not have donor disclosure requirements. However, advertisers that only run issue ads -- meaning ads that do not refer to candidates or urge voters how to vote -- may not register with the FEC because they are exempt from regulation under friendly Supreme Court jurisprudence around issue speech. Any legislative or regulatory reforms requiring disclosure for online political ads would be unlikely to reach these groups. FEC-registered advertisers accounted for 38\% of all political ads and 3\% of advertisers in Facebook's ad library, although the library includes many promotions, surveys, and ads from state and local political groups not regulated by the FEC. An audit of Facebook's political ad library estimated that 55\% of ads involved non-political content, but that the library included 99.8\% of ads from FEC-registered advertisers~\citep{le2022audit}. Furthermore, in order to test our hypothesis on funding transparency, we limit our scope of outside groups to advertisers the CRP identified. We include all FEC-registered candidates and parties, since they are required to disclose donors. Even with these coding decisions, we retained 85\% of the ads from FEC-registered advertisers.

\subsection{Advertiser Accountability}

We compare advertisers based on two different avenues for accountability: (1) donor disclosure, and (2) group persistence. Advertisers face different donor disclosure requirements based on the content of the speech and the speaker~\citep{wood_arlss}. For outside groups, an important distinction arises from whether they only run independent expenditures, which by definition are not coordinated with campaigns. We summarize donor disclosure requirements by advertiser type below: 

\begin{itemize}
    \item \textbf{Candidates and Parties}: disclose all donors over \$200\footnote{For all advertiser types, minimums for each donor are aggregated across donations and within a single election.}
    \item \textbf{Traditional PACs}: can coordinate with candidates;\\disclose all donors over \$200, with contribution limits
    \item \textbf{Super PACs}: only independent expenditures;\\ disclose all donors over \$200, no contribution limits
    \item \textbf{Hybrid PACs}: have two bank accounts, one that acts like a Traditional PAC and one that acts like a Super PAC;\\ disclose all donors over \$200 to either account
    \item \textbf{501(c) Organizations and Corporations}: no donor \\disclosure required and no contribution limits
\end{itemize}
The CRP categorizes outside groups in three bins, by the percentage of disclosed donor contributions that they report: full ($>$95\%), partial\footnote{Partial disclosure (``gray money'') often occurs when a non-disclosing group gives to a disclosing group, so that the ``money trail'' ends with the non-disclosing group name, and no specific donor can be identified. We code these as ``disclosed'' to (conservatively) bias our estimates against our hypothesis, but classifying them as ``dark money'' does not meaningfully change our results (see Appendix Table~\ref{appendix:gray_money}).}  (5\%-95\%), or none ($<$5\%). Using this information, we classify outside groups by donor transparency as follows:

\begin{itemize}

\item \textbf{Dark Money}: $<$5\% of contributions disclosed
\item \textbf{Disclosed}: $\geq$5\% of contributions disclosed
\end{itemize}
To analyze advertiser permanence, we cross-referenced FEC-registration in the 2016, 2018, and 2020 election cycles. We then distinguished outside groups by their permanence as follows: 
\begin{itemize}

\item \textbf{Disappeared}: registered w/FEC only in 2018
\item \textbf{Persistent}: registered w/FEC in multiple cycles
\end{itemize}
We use the above definition of ``disappearing'' as a proxy for whether advertisers ran ads in only one cycle, since we could not collect ads for other elections. Since candidates and parties are required to disclose donors and cannot disappear, we group them with disclosed and persistent outside groups in our analysis\footnote{Appendix Table~\ref{appendix:outside_groups} shows results for outside groups only.}.  

\subsection{Sentiment Computation} Our dependent variable is ad sentiment, which we primarily measured using the \textbf{Valence Aware Dictionary and Sentiment Reasoner (VADER)}~\citep{vader}. VADER is a lexicon and rule-based sentiment analysis tool specifically attuned to sentiments expressed on social media. Several prior works have used VADER for analyzing the sentiment of political content on social media~\citep{dutt2019senator, hussain2021artificial}, as well as for other tasks such as hate speech and fake news detection~\citep{davidson2017hateSpeech, shu2020fakenewsnet}. Following other studies of Facebook's ad library~\citep{edelson_transparency, fowler_apsr}, we consider each ad record\footnote{Online ads are often set to run for a certain number of impressions or targeting criteria. When campaigns see that ads are successful, they will renew them, making them appear in Facebook's ad library as another record. Since each of these represents additional impressions, we include them to gauge sentiment as a whole. Removing duplicates does not meaningfully affect our results (see Appendix Table~\ref{appendix:duplicates}).} in the library as our unit of analysis. Using VADER, we computed the overall sentiment of each ad record on a scale of -1 (extreme negative) to +1 (extreme positive). Appendix Table~\ref{appendix:ads} includes examples of ads and their sentiment scores. This work limits sentiment analysis to English ad text; we do not consider other languages, ad images or videos. 

We validate our VADER sentiment scores through large language models (LLMs) and Latent Dirichlet allocation (LDA) topic modeling. Specifically, we use LLMs to classify ads by polarity, stance, and emotions using the following fine-tuned versions of the RoBERTa model\footnote{We choose these because they are among the most downloaded open-source models on Hugging Face for sentiment-related tasks.}~\citep{liu2019roberta}:

\begin{itemize}
    \item \textbf{Polarity}: positive, negative, neutral\\~\citep{barbieri-etal-2020-tweeteval}
    \item \textbf{Stance} (topic-agnostic): for, against, neutral\\~\citep{gajewska-2023-eevvgg}
    \item \textbf{Emotions}: anger, disgust, fear, joy, sadness, surprise, neutral~\citep{hartmann2022emotionenglish}
\end{itemize}
Each model outputs class probabilities that sum to 1 (e.g. for a given ad, the polarity model could output 0.6 for positive, 0.1 for negative, and 0.3 for neutral). As expected, the VADER sentiment score was negatively correlated (Pearson correlation coefficient $r$) with the probability of ``negative'' polarity (-0.46) and the stance of ``against'' (-0.31). The emotions of anger (-0.20), disgust (-0.21), fear (-0.19), and sadness (-0.21) were also negatively correlated with VADER, while joy (0.27) and surprise (0.05) were positively correlated. Additionally, we use LDA topic modeling to compare ads in the upper and lower quartile of VADER sentiment (see Appendix Tables~\ref{appendix:topics_negative} and \ref{appendix:topics_positive}). In our main analysis, we primarily present results with VADER given the focus of this work on sentiment and its precedent in previous work, and supplement with the LLM measures and LDA analysis to demonstrate the robustness of our results. 

\begin{figure*}[t!] 
    \centering
    \includegraphics[width=\textwidth]{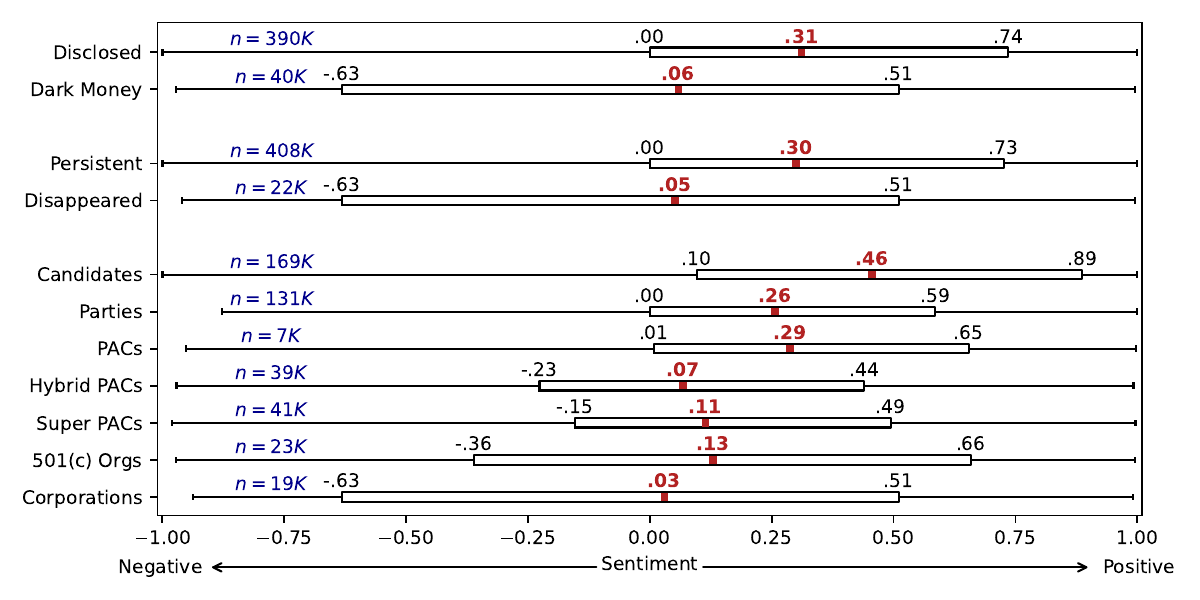}
    \caption{\textbf{Average Sentiment and Interquartile Range by Advertiser Type}. The distribution of VADER sentiment scores is more negative for advertiser types with less accountability (by donor disclosure, group persistence, and categories of outside groups). The red center lines denote means, the box extends from the first quartile to the third quartile (IQR), and the whiskers extend from the box to 1.5x the IQR. The number of advertisements ($n$) is reported in blue on top of each category's left whisker.}
    \label{fig:sentiment_boxplot}
\end{figure*}
\begin{table*}[t!]
\small
\centering
\caption{Estimated Differences in Sentiment by Advertiser Accountability}
\label{tab:reg_coeff}
\begin{tabular}{lccc|cc|ccc} 
\multicolumn{9}{c}{(1, 6-8) All ads (2) All ads with fixed effects for advertiser type (3) Only ads from 501(c) organizations and corporations} \\[0.5mm]
\multicolumn{9}{c}{(4) Only advertisers with a Democratic or liberal viewpoint (5) Only advertisers with a Republican or conservative viewpoint } \\[1mm]

\toprule
& \makecell{(1)} & \makecell{(2)} & \makecell{(3)} & \makecell{(4)} & \makecell{(5)} & \makecell{(6)} & \makecell{(7)} & \makecell{(8)} \\[1mm]
\toprule
\multirow{2}{2.2cm}{Dark Money\\\& Disappeared} & $-$0.290*** & $-$0.265*** & $-$0.370*** & $-$0.231* & $-$0.235* & 0.213** & 0.430** & 0.495** \\
& (0.095) & (0.099) & (0.087) & (0.131) & (0.142) & (0.105) & (0.172) & (0.200) \\[2mm]
\multirow{2}{2.2cm}{Dark Money\\\& Persistent} & $-$0.212* & $-$0.221** & $-$0.288** & $-$0.214* & $-$0.209* & 0.120* & $-$0.046 & 0.158*** \\ 
& (0.127) & (0.100) & (0.126) & (0.110) & (0.108) & (0.073) & (0.063) & (0.039) \\[2mm]
\multirow{2}{2.2cm}{Disclosed\\\& Disappeared} & 0.182* & 0.235** & 0.109 & 0.155 & 0.580*** & $-$0.102** & $-$0.116** & 0.082* \\ 
& (0.098) & (0.105) & (0.091) & (0.096) & (0.185) & (0.050) & (0.048) & (0.045) \\[1mm]
\midrule
\multirow{2}{2.2cm}{\makecell[l]{Intercept (Disclosed\\\& Persistent)}} &  0.310*** & 0.456*** & 0.390*** & 0.366*** & 0.583*** & 0.245*** & 0.252*** & 0.079*** \\ 
& (0.060) & (0.077) & (0.045) & (0.043) & (0.092) & (0.033) & (0.026) & (0.012) \\[1mm]
\midrule
Sentiment Measure & \multirow{2}{*}{VADER} & \multirow{2}{*}{VADER} & \multirow{2}{*}{VADER} & \multirow{2}{*}{VADER} & \multirow{2}{*}{VADER} & \multicolumn{3}{l}{LLM Polarity\hspace{5pt}LLM Stance\hspace{7pt}LLM Emotion} \\
(Outcome Variable) & &&&&& (``Negative'') & (``Against'') & (``Fear'') \\[2mm]
Number of Ads &  430,044 & 430,044 & 46,959 & 232,129 & 193,686 & 430,044 & 430,044 & 430,044 \\[2mm]

\multicolumn{2}{l}{Adv Type Fixed Effects\hspace{10pt}No} & Yes & N/A & Yes & Yes & No & No & No \\

\bottomrule\\
\multicolumn{9}{l}{Regression coefficients represent estimated differences in sentiment for each level of advertiser accountability, compared to the intercept}\\
\multicolumn{9}{l}{term of fully-transparent groups (disclosed \& persistent). Standard errors are clustered at the advertiser-level and reported in parenthesis}\\
\multicolumn{9}{l}{below each coefficient. Statistically significant differences are denoted by $*$ $p <$ 0.1,  $*$$*$ $p <$ 0.05, and $*$$*$$*$ $p <$ 0.01, based on two-sided}\\
\multicolumn{9}{l}{t-tests for coefficients $\neq$ 0. The intercept for (2) \& (4)-(5) includes candidates as the omitted fixed effect for advertiser types.}\\[2mm]

\multicolumn{9}{l}{We expected more negative ads with less advertiser accountability (dark money, disappeared, or both). The VADER sentiment measure}\\
\multicolumn{9}{l}{ is on a scale of -1 (extreme negative) to +1 (extreme positive), so we expect to see negative coefficients for key explanatory variables in}\\
\multicolumn{9}{l}{Models (1)-(5). The LLM measures are class probabilities on a scale of 0 to 1, so we expect positive coefficients in Models (6)-(8).}\\

\end{tabular} 
\end{table*}

\section{Analysis}

The most negative advertising came from dark money outside groups that also disappeared after the election. This interaction of dark money and impermanence was also the most significantly associated with a decrease in ad sentiment ($p=0.002$). However, dark money alone was also associated with a significant decrease in ad sentiment ($p=0.093$), while impermanence alone was not.

\subsection{Sentiment by Advertiser Accountability} Figure~\ref{fig:sentiment_boxplot} reports the mean and interquartile range of ad sentiment as well as the number of ads run by different advertiser types\footnote{Appendix Table~\ref{appendix:summarystats} also reports the number of ads and advertisers for each advertiser type.}. On the scale of -1 (extreme negative) to +1 (extreme positive), dark money ads had an average sentiment of 0.06 compared to 0.31 for ads with disclosed donors. Disappearing advertisers similarly ran ads with lower sentiment on average than persistent groups (0.05 to 0.30). The overall interquartile range of ad sentiment was entirely on the positive side ([0.00 to 0.71]). However, nearly 40\% of ads funded by dark money or run by disappearing groups had negative sentiment (i.e. sentiment $<$ 0). Dark money or disappearing advertisers were twice as likely to run negative ads than disclosed or persistent groups, respectively. 

Advertiser types with avenues for less accountability also had more negative sentiment. Outside groups overall had an average sentiment of 0.10 compared to 0.46 for candidates and 0.26 for parties. Among outside groups, traditional PACs ran ads with higher sentiment on average than all other groups (0.29 to 0.09). In particular, 501(c) organizations and corporations -- advertiser types created after \textit{Citizens United} -- had an average sentiment of 0.13 and 0.03, respectively. These results suggest that more negative advertising comes from advertisers with less accountability, but it is unclear whether the key statistical associations in the data are with donor disclosure, group permanence, or advertiser type.

\subsection{Dark Money Is Significantly Associated With More Negative Sentiment} We use linear regression models to determine if there is a statistically significant difference in advertising sentiment related to dark money, advertiser impermanence, or both. We cluster all standard errors at the advertiser-level given that we categorize advertisers by accountability and not individual ads. Table~\ref{tab:reg_coeff} shows the regression coefficients which represent the estimated differences in sentiment by advertiser accountability for several different model specifications. Models (1)-(5) report coefficients using VADER for different subsets of advertisers: (1) all ads, (2) all ads with fixed effects for advertiser type, (3) only ads from 501(c) organizations and corporations, (4) only advertisers with a Democratic or liberal viewpoint, and (5) only advertisers with a Republican or conservative viewpoint.  

Across all specifications using VADER, dark money is significantly associated a decrease in ad sentiment ($p<0.1$) regardless of advertiser persistence. However, this association is strongest for dark money advertisers that also disappeared after the election. We perform F-tests between our model coefficients to determine if there are statistically significant differences\footnote{Since the F-test is not directional and corresponds to a two-sided t-test, we take $p<0.1$ as our significance level.} in sentiment by accountability. Table~\ref{tab:f-test} reports the F-test results with Model (1) for interpretability of the intercept term, but the significant coefficient pairs are the same with fixed effects for advertiser type and for other model specifications with VADER (see Appendix Tables~\ref{appendix:significance2}-\ref{appendix:significance8}). When compared to disclosed and persistent advertisers, dark money alone is associated with a drop in sentiment of -0.21 ($p=0.093$), while dark money ads from disappearing groups related to a -0.29 decrease in sentiment ($p=0.002$). But when fixing the level of donor disclosure, there is no statistically significant difference in sentiment due to a change in advertiser impermanence. 

\begin{figure}[t!] 
    \centering
    \includegraphics[width=\linewidth]{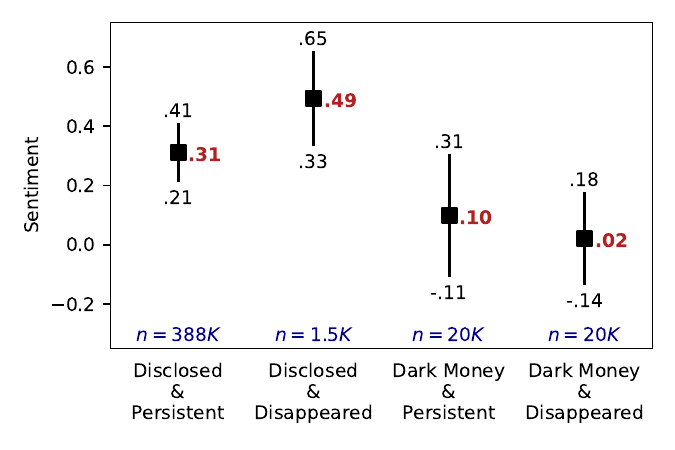}
    \caption{Estimated Sentiment (90\% CI) by Accountability\\[1mm] \centering{\small{Based on Coefficients in Table 1 - Model (1)}}}
    \label{fig:reg_coeff}
\end{figure}

Figure~\ref{fig:reg_coeff} illustrates these results by showing the estimated levels of sentiment by our two pathways of accountability using the results of Model (1), with 90\% confidence intervals. Dark money and disappearing advertisers had an estimated sentiment of [-0.14, 0.18], compared to [0.21, 0.41] for disclosed and persistent advertisers. The sentiment for disclosed and disappearing groups appears to be the highest but we caution that there were only 1,487 ads in this category. These results also hold when separately analyzing\footnote{When comparing political-viewpoints, we include fixed effects for advertiser type to account for incumbency and fundraising differences. Recall that outside groups who make independent expenditures only -- SuperPACs, corporations, and 501(c) organizations -- are not technically party affiliated.} ads run by Democrats (or liberal groups) and Republicans (or conservative groups) (Models (4) and (5)), with nearly identical coefficient estimates for the decrease associated with dark money. 

\subsection{Sentiment of 501(c) Organizations and Corporations} While many Hybrid and Super PACs in our data have limited disclosure, every (fully) dark money \textit{and} disappearing advertiser in our dataset was a 501(c) organization or corporation. As we discussed above, 501(c) organizations and corporations emerged as frequent advertiser types after \textit{Citizens United} and do not have any donor disclosure requirements in federal elections. The Internal Revenue Code section 501(c)\footnote{Section 501(c)4 ``social welfare'' organizations are the most prevalent 501(c) organization type spending in recent elections.} lists various types of non-profit organizations, some of which can spend in political campaigns. We observe that 501(c) organizations and corporations differ on persistence: 94\% of corporations were funded by dark money, and nearly all of these disappeared after one election. Meanwhile, 92\% of 501(c) organizations were funded by dark money but only 11\% disappeared after one election. Since any transparency by these groups is voluntary, we replicate our regression analysis only among ads from 501(c) organizations and corporations (Table~\ref{tab:reg_coeff} - Model (3)). Among these advertiser types, we again find that dark money is associated with a statistically significant decrease in ad sentiment regardless of group permanence. However, when we look only among corporations and 501(c) organizations, we observe larger estimated differences in sentiment. Dark money alone is associated with a decrease in sentiment of -0.29 ($p=4.7\mathrm{e}{-6}$), while dark money and impermanence are associated with a decrease of -0.37 ($p=7.8\mathrm{e}{-11}$).

\subsection{Association With LDA Topics and LLM Measures of Polarity, Stance, and Emotion} For dark money and disappearing advertisers, we similarly find a strong association ($p<0.05$) with the LLM measures of negative polarity, negative stance, and the emotion of fear. Models (6)-(8) in Table~\ref{tab:reg_coeff} report the estimated differences in these LLM measures by advertiser accountability. In contrast to VADER, we expect to see a positive association with less transparency because the LLM measures are class probabilities on a scale of 0 to 1. For negative polarity, Model (6) shows that dark money alone is associated with an increase of 0.12 ($p=0.043$) when compared to disclosed and persistent advertisers, while the interaction of dark money and disappearing related to an increase of 0.21 ($p=0.100$). These results are very similar to Model (1) with VADER scores, indicating a robustness to the choice of sentiment measure. 

For negative stances, Model (7) shows that only dark and disappearing groups have a positive association ($p=0.012$). However, these ads were nearly 3 times more likely to be classified as having a negative stance than ads from disclosed and persistent advertisers (0.68 to 0.25). Appendix Table~\ref{appendix:emotion} shows the model coefficients for advertiser accountability and each of the emotion classes (anger, disgust, fear, joy, sadness, surprise, neutral). Across all these classes of emotion, only fear had a significant positive association ($p<0.1$) with less advertiser accountability (Model (8) in Table~\ref{tab:reg_coeff}). When compared to ads from disclosed and persistent advertisers, ads from dark and disappearing advertisers were over 7 times more likely to be classified with the emotion of fear (0.57 to 0.08, $p=0.013$). Even ads from dark money and persistent advertisers were 3 times more likely to be classified with fear (0.24 to 0.08, $p=0.071$). 

Based on the LDA topic analysis, positive ads involve more topics related to increasing support for campaigns, while negative ads involve more topics related to attacking other campaigns. Appendix Tables~\ref{appendix:topics_negative} and \ref{appendix:topics_positive} show the top 10 topics and their word frequencies for ads in the lower and upper quartile of VADER sentiment, respectively. Positive ads (those in the upper quartile) have topics with more words associated with increasing the support for campaigns, such as ``join'', ``help'', and ``support''. In contrast, negative ads (those in the lower quartile) have more words associated with attacking other campaigns, such as ``fight'', ``attack'', and ``bad''. However, many topics are associated with both positive and negative ads, such as voter turnout, fundraising, healthcare, and Brett Kavanaugh. Other specific topics such as tax cuts, election fraud, guns, and climate change only show up in negative ads; such relative policy specificity may relate to the possibility that negative ads can be used to inform (regardless of their factual accuracy)~\citep{dowling_negative}. These results are intended to serve as a robustness check of our analysis of VADER sentiment. Future researchers may want to explore the relationship between sentiment and certain topics, especially as it relates to advertiser accountability. 



\section{Discussion}

\begin{table}[t!]
\small
\centering
\caption{Significant Diff. in Sentiment by Accountability}
\label{tab:f-test}
\begin{tabular}{lccc} 
\toprule
\makecell[l]{F-Statistic\\(P-Value)} & \makecell{Disclosed\\\&\\Persistent} & \makecell{Disclosed\\\&\\Disappeared} & \makecell{Dark Money\\\&\\Persistent} \\[1mm]
\toprule
\multirow{2}{2.2cm}{Dark Money\\\& Disappeared} & 9.31*** & 19.62*** & 0.35 \\
& (0.002) & ($9.5\mathrm{e}{-6}$) & (0.557) \\[2mm]
\multirow{2}{2.2cm}{Dark Money\\\& Persistent} & 2.82* & 8.50***\\ 
& (0.093) & (0.004) \\[2mm]
\multirow{2}{2.2cm}{Disclosed\\\& Disappeared} & 3.48*  \\ 
& (0.062) \\
\bottomrule\\
\multicolumn{4}{c}{F-Test between coefficients in Table~\ref{tab:reg_coeff} - Model (1), with }\\
\multicolumn{4}{c}{p-values reported in parentheses below each F-statistic.}\\
\multicolumn{4}{c}{$*$ $p <$ 0.1  $*$$*$ $p <$ 0.05  $*$$*$$*$ $p <$ 0.01} \\
\end{tabular} 
\end{table}

As online political advertising continues to grow, policymakers need guidance on adapting campaign finance laws. This work provides an empirical analysis of how dark money involvement and group permanence affect the tone of Facebook political ads from the 2018 election. Political advertisers on Facebook produce ads with significantly different sentiment depending on whether they disclose donors. Among FEC-registered political committees, the most negative advertising came from dark money outside groups, particularly if they also disappeared and only registered with the FEC for the 2018 cycle. 

While previous studies about television and radio ads have also found outside groups to be the most negative, the prevalence of outside groups on Facebook represents a stark difference from offline ads where candidates and parties still dominate the airwaves~\citep{fowler_negative, fowler_apsr}. Moreover, TV and radio ads are more tightly regulated than online platforms. The larger volume and negativity of dark money ads on Facebook suggests that stricter campaign finance disclosure laws on social media may affect the composition and tone of online advertisements. Whatever the informational benefits or polarization detriments of negative ads, increasing donor transparency could reduce the negativity of online political advertising.

\subsection{Limitations} We caution that our findings adopt some of the limitations of Facebook's ad library during the 2018 election, in addition to other constraints of our scope and analysis. First, we are beholden to Facebook's definition of what constitutes a political ad. As we explained in our methods above, we consider each ad record as our unit of analysis, but without granular impressions or targeting data, we could not weight ads by their influence on voters. For example, a single negative ad shown nationwide would likely affect the overall tone of content viewed by voters more than multiple ads shown to smaller groups of people. Second, we rely on the ``Paid for by'' label associated with each ad to determine advertiser names and types. But during the 2018 election, Facebook did not prevent advertisers from providing inconsistent, deceptive, or even missing advertiser names. We therefore may have incorrectly determined some advertiser types, despite manually verifying our matches of advertiser names across Facebook, FEC data, and other sources. 

Furthermore, we limit our analysis to speech from FEC-registered advertisers. A large quantity of political speech on social media is unpaid, and we only analyze paid speech. While our scope allows us to draw conclusions about the sentiment of advertisers subject to federal regulation, we cannot generalize to the overall tone of political speech online. Future research may want to compare content from other accounts to our analysis of FEC-registered advertisers. Finally, our research design and findings are descriptive, not causal. All advertisers selected into their level of accountability (donor disclosure and permanence). Our analysis concludes that dark money is more associated with negativity than permanence, but this does not speak to causality.  

\subsection{Future Work} In addition to donor disclosure and group permanence, ad targeting and delivery represent further accountability deficits of online political advertising. Social media platforms such as Facebook enable online advertisers to target small groups of users based on personal information, demographics, or other interests inferred from online behavior. This type of ``microtargeting'' makes it easier to single out susceptible voters with disinformation or speech that triggers their underlying fears and suspicions. If the microtargeted group is homogeneous enough, it is unlikely that any viewer will ``blow the whistle'' on disinformation or other harmful speech, making microtargeting a potential avenue for speech without accountability. Moreover, ad delivery algorithms can also play a crucial role in which users ultimately see a given ad. Facebook may choose to optimize ad delivery based on user engagement, which can further skew impressions along demographic groups beyond the control of the advertiser~\citep{ali2021ad}. While Facebook releases limited targeting criteria\footnote{Facebook started releasing this data after we conducted this study, and it is only available for the past 90 days.} at the \textit{advertiser} level in its political ad library, the targeting criteria used for specific ads is not available. Furthermore, Facebook only provides high-level demographic data on user impressions, which yields little transparency on how ad delivery algorithms affect the relationship between targets and reach. Publishing targeting and delivery information about individual ads would enable scholars to study whether this accountability deficit also facilitates more negative advertising, in the same way as we show for campaign finance transparency. If Facebook will not release the detailed targeting criteria for each ad, government regulators could mandate transparency of targeting information, which is available for broadcast and other media sources~\citep{wood_sclr}.

Research on Facebook political ads is just beginning, and opportunities for future work abound. We study the sentiment of ad text, and others have studied mentions of opponents~\citep{fowler_apsr} as well as stance detection~\citep{islam2023microtargeting}, but online ads also contain videos and images to be analyzed. Cross-platform studies may provide additional leverage: Google has required political advertisers to register with the FEC, allowing future researchers to study the effects of stricter requirements by online platforms. While Facebook does not release microtargeting criteria in its ad library, other platforms may allow researchers to study this additional pathway for avoiding accountability. Finally, this is a study of advertising during one midterm election. Repeating it would help teach us how incumbency and party control interact with accountability over time.

\bibliography{ref}

\clearpage
\onecolumn
\section{Appendix}

\begin{itemize}
\setlength\itemsep{2mm}

\item \textbf{Table~\ref{appendix:summarystats}: Summary Statistics} - Number of ads and advertisers for different advertiser types.

\item \textbf{Tables~\ref{appendix:significance2}-\ref{appendix:significance8}: F-test significance results between coefficients for Models (2)-(8) from Table~\ref{tab:reg_coeff}} in our main analysis. Across all model specifications with VADER, the significant coefficient pairs are the same as for Model (1) in Table~\ref{tab:f-test} at the level of $p<0.1$ (i.e. only dark money is associated with a significant difference in sentiment, regardless of advertiser persistence).

\item \textbf{Table~\ref{appendix:duplicates}: Regression coefficients with duplicate ad records removed}. Online ads are often set to run for a certain number of impressions or targeting criteria. When campaigns see that ads are successful, they will renew them, making them appear in Facebook's ad library as another record. Removing duplicates does not change the result that dark money is significantly associated with more negative sentiment. Similar to Table~\ref{tab:reg_coeff}, coefficients represent the estimated change in sentiment by accountability.   

\item \textbf{Table~\ref{appendix:gray_money}: Regression coefficients with partial disclosure coded as dark money}. The CRP categorizes outside groups under partial disclosure if they disclose anywhere between 5\% to 95\% of their donor contributions. Partial disclosure (``gray money'') often occurs when a non-disclosing group gives to a disclosing group, so that the ``money trail'' ends with the non-disclosing group name, and no specific donor can be identified. We code these as ``disclosed'' in our main analysis to (conservatively) bias our estimates against our hypothesis, but classifying them as ``dark money'' shows that any decrease in donor disclosure (i.e. dark or gray money) is significantly associated with more negative sentiment. Similar to Table~\ref{tab:reg_coeff}, coefficients represent the estimated change in sentiment by accountability. 

\item \textbf{Table~\ref{appendix:outside_groups}: Regression coefficients with ads from outside groups only}. Since candidates and parties are required to disclose donors and cannot disappear, we them with disclosed and persistent outside groups for our main analysis. Among only outside groups, dark money is still significantly associated with more negative sentiment. Similar to Table~\ref{tab:reg_coeff}, coefficients represent the estimated change in sentiment by accountability.    

\item \textbf{Tables~\ref{appendix:topics_negative}-\ref{appendix:topics_positive}: Latent Dirichlet allocation (LDA) topic analysis}. We report the top 10 LDA topics and their word frequencies for ads in the upper and lower quartile of VADER sentiment. These results are intended to serve as a robustness check of our analysis with VADER, and future work may want to further explore the relationship between sentiment and certain topics, especially as it relates to advertiser accountability. 

\item \textbf{Table~\ref{appendix:ads}: Examples of ads and sentiment scores} from VADER between -1 (extreme negative) to +1 (extreme positive).

\end{itemize}

\begin{table}[ht]
\small
\centering
\caption{Number of Ads and Advertisers by Advertiser Type}
\label{appendix:summarystats}
\begin{tabular}{lcccc} 
\toprule
Advertiser Type & \# Ads & \% Ads & \# Advertisers & \% Advertisers \\
\toprule
Disclosed \& Persistent & 388,161 & 90.3\% & 1,171 & 91.7\% \\
Disclosed \& Disappeared & 1,487 & 0.3\% & 30 & 2.3\% \\
Dark Money \& Persistent & 20,017 & 4.7\% & 38 & 3.0\% \\
Dark Money \& Disappeared & 20,379 & 4.7\% & 38 & 3.0\% \\
\midrule
Disclosed & 389,648 & 90.6\% & 1,201 & 94.0\% \\
Dark Money & 40,396 & 9.4\% & 76 & 6.0\% \\
\midrule
Persistent & 408,178 & 94.9\% & 1,209 & 94.7\% \\
Disappeared & 21,866 & 5.1\% &  68 & 5.3\% \\
\midrule
Candidates & 168,849 & 39.3\%  & 884 & 69.2\% \\
Parties & 131,260 & 30.5\%  & 101 & 7.9\%  \\
PACs & 7,093 & 1.6\%  & 36 & 2.8\% \\
Hybrid PACs & 38,918 & 9.0\% & 25 & 2.0\% \\
Super PACs & 41,453 & 9.6\% & 125 & 9.8\% \\
501(c) Orgs & 23,435 & 5.4\% & 83 & 6.5\% \\
Corporations & 19,036 & 4.4\% & 23 & 1.8\% \\
\midrule
Democratic/Liberal & 232,129 & 54.0\% &  675 & 57.2\% \\
Republican/Conservative & 193,686 & 46.0\% & 505 & 42.8\% \\
\midrule
All & 430,044 & - & 1,277 & - \\
\bottomrule\\
\end{tabular} 
\end{table}
\begin{table}[h]
\small
\centering
\caption{Significant Differences by Accountability\\[1mm]\centering{\textbf{For Model (2) With Adv Type Fixed Effects}}}
\label{appendix:significance2}
\begin{tabular}{lccc} 
\toprule
\makecell[l]{F-Statistic\\(P-Value)} & \makecell{Disclosed\\\&\\Persistent} & \makecell{Disclosed\\\&\\Disappeared} & \makecell{Dark Money\\\&\\Persistent} \\[1mm]
\toprule
\multirow{2}{2.2cm}{Dark Money\\\& Disappeared} & 7.16*** & 17.35*** & 0.09\\
& (0.007) & ($3.1\mathrm{e}{-5}$) & (0.758) \\[2mm]
\multirow{2}{2.2cm}{Dark Money\\\& Persistent} & 4.86** & 11.60*** \\
& (0.027) & ($6.6\mathrm{e}{-4}$)\\[2mm]
\multirow{2}{2.2cm}{Disclosed\\\& Disappeared} &  4.99**  \\ 
& (0.025) \\
\bottomrule\\
\multicolumn{4}{l}{F-Test between coefficients in Table~\ref{tab:reg_coeff} - Model (2), with }\\
\multicolumn{4}{l}{p-values reported in parentheses below each F-statistic.}\\
\multicolumn{4}{l}{$*$ $p <$ 0.1  $*$$*$ $p <$ 0.05  $*$$*$$*$ $p <$ 0.01} \\
\end{tabular} 
\end{table}

\begin{table}[h]
\small
\centering
\caption{Significant Differences by Accountability\\[1mm]\centering{\textbf{For Model (3) With Only 501(c) Orgs and Corporations}}}
\label{appendix:significance3}
\begin{tabular}{lccc} 
\toprule
\makecell[l]{F-Statistic\\(P-Value)} & \makecell{Disclosed\\\&\\Persistent} & \makecell{Disclosed\\\&\\Disappeared} & \makecell{Dark Money\\\&\\Persistent} \\[1mm]
\toprule
\multirow{2}{2.2cm}{Dark Money\\\& Disappeared} & 18.12*** & 19.51* & 0.35 \\
& ($2.1\mathrm{e}{-5}$) & ($1.0\mathrm{e}{-5}$) & (0.553) \\[2mm]
\multirow{2}{2.2cm}{Dark Money\\\& Persistent} & 5.24** & 7.86* \\
& (0.022) & (0.005) \\[2mm]
\multirow{2}{2.2cm}{Disclosed\\\& Disappeared} & 1.43  \\ 
& (0.232) \\
\bottomrule\\
\multicolumn{4}{l}{F-Test between coefficients in Table~\ref{tab:reg_coeff} - Model (3), with }\\
\multicolumn{4}{l}{p-values reported in parentheses below each F-statistic.}\\
\multicolumn{4}{l}{$*$ $p <$ 0.1  $*$$*$ $p <$ 0.05  $*$$*$$*$ $p <$ 0.01} \\
\end{tabular} 
\end{table}
\begin{table}[h]
\small
\centering
\caption{Significant Differences by Accountability\\[1mm]\centering{\textbf{For Model (4) With Democratic/Liberal Advertisers}}}
\label{appendix:significance4}
\begin{tabular}{lccc} 
\toprule
\makecell[l]{F-Statistic\\(P-Value)} & \makecell{Disclosed\\\&\\Persistent} & \makecell{Disclosed\\\&\\Disappeared} & \makecell{Dark Money\\\&\\Persistent} \\[1mm]
\toprule
\multirow{2}{2.2cm}{Dark Money\\\& Disappeared} & 3.09* & 7.02*** & 0.01\\
& (0.079) & (0.008) & (0.93) \\[2mm]
\multirow{2}{2.2cm}{Dark Money\\\& Persistent} & 3.79* & 6.59** \\
& (0.052) & (0.010)\\[2mm]
\multirow{2}{2.2cm}{Disclosed\\\& Disappeared} &  2.62  \\ 
& (0.105) \\
\bottomrule\\
\multicolumn{4}{l}{F-Test between coefficients in Table~\ref{tab:reg_coeff} - Model (4), with }\\
\multicolumn{4}{l}{p-values reported in parentheses below each F-statistic.}\\
\multicolumn{4}{l}{$*$ $p <$ 0.1  $*$$*$ $p <$ 0.05  $*$$*$$*$ $p <$ 0.01} \\
\end{tabular} 
\end{table}
\begin{table}[h]
\small
\centering
\caption{Significant Differences by Accountability\\[1mm]\centering{\textbf{For Model (5) With Republican/Conservative Advertisers}}}
\label{appendix:significance5}
\begin{tabular}{lccc} 
\toprule
\makecell[l]{F-Statistic\\(P-Value)} & \makecell{Disclosed\\\&\\Persistent} & \makecell{Disclosed\\\&\\Disappeared} & \makecell{Dark Money\\\&\\Persistent} \\[1mm]
\toprule
\multirow{2}{2.2cm}{Dark Money\\\& Disappeared} & 2.72* & 16.47*** & 0.06\\
& (0.099) & ($4.9\mathrm{e}{-5}$) & (0.810) \\[2mm]
\multirow{2}{2.2cm}{Dark Money\\\& Persistent} & 3.73* & 18.51*** \\
& (0.053 & ($1.7\mathrm{e}{-5}$)\\[2mm]
\multirow{2}{2.2cm}{Disclosed\\\& Disappeared} &  9.81***  \\ 
& (0.002) \\
\bottomrule\\
\multicolumn{4}{l}{F-Test between coefficients in Table~\ref{tab:reg_coeff} - Model (5), with }\\
\multicolumn{4}{l}{p-values reported in parentheses below each F-statistic.}\\
\multicolumn{4}{l}{$*$ $p <$ 0.1  $*$$*$ $p <$ 0.05  $*$$*$$*$ $p <$ 0.01} \\
\end{tabular} 
\end{table}
\begin{table}[h]
\small
\centering
\caption{Significant Differences by Accountability\\[1mm]\centering{\textbf{For Model (6) Using LLM Polarity (``Negative'' Class)}}}
\label{appendix:significance6}
\begin{tabular}{lccc} 
\toprule
\makecell[l]{F-Statistic\\(P-Value)} & \makecell{Disclosed\\\&\\Persistent} & \makecell{Disclosed\\\&\\Disappeared} & \makecell{Dark Money\\\&\\Persistent} \\[1mm]
\toprule
\multirow{2}{2.2cm}{Dark Money\\\& Disappeared} & 4.09** & 8.74*** & 0.60\\
& (0.043) & (0.003) & (0.437) \\[2mm]
\multirow{2}{2.2cm}{Dark Money\\\& Persistent} & 2.70* & 8.74*** \\
& (0.100) & (0.003)\\[2mm]
\multirow{2}{2.2cm}{Disclosed\\\& Disappeared} &  4.12**  \\ 
& (0.042) \\
\bottomrule\\
\multicolumn{4}{l}{F-Test between coefficients in Table~\ref{tab:reg_coeff} - Model (6), with }\\
\multicolumn{4}{l}{p-values reported in parentheses below each F-statistic.}\\
\multicolumn{4}{l}{$*$ $p <$ 0.1  $*$$*$ $p <$ 0.05  $*$$*$$*$ $p <$ 0.01} \\
\end{tabular} 
\end{table}
\begin{table}[h]
\small
\centering
\caption{Significant Differences by Accountability\\[1mm]\centering{\textbf{For Model (7) Using LLM Stance (``Against'' Class)}}}
\label{appendix:significance7}
\begin{tabular}{lccc} 
\toprule
\makecell[l]{F-Statistic\\(P-Value)} & \makecell{Disclosed\\\&\\Persistent} & \makecell{Disclosed\\\&\\Disappeared} & \makecell{Dark Money\\\&\\Persistent} \\[1mm]
\toprule
\multirow{2}{2.2cm}{Dark Money\\\& Disappeared} & 6.26** & 9.76*** & 7.05***\\
& (0.012) & (0.002) & (0.008) \\[2mm]
\multirow{2}{2.2cm}{Dark Money\\\& Persistent} & 0.54 & 0.97 \\
& (0.461) & (0.326)\\[2mm]
\multirow{2}{2.2cm}{Disclosed\\\& Disappeared} &  5.83**  \\ 
& (0.016) \\
\bottomrule\\
\multicolumn{4}{l}{F-Test between coefficients in Table~\ref{tab:reg_coeff} - Model (7), with }\\
\multicolumn{4}{l}{p-values reported in parentheses below each F-statistic.}\\
\multicolumn{4}{l}{$*$ $p <$ 0.1  $*$$*$ $p <$ 0.05  $*$$*$$*$ $p <$ 0.01} \\
\end{tabular} 
\end{table}
\begin{table}[h]
\small
\centering
\caption{Significant Differences by Accountability\\[1mm]\centering{\textbf{For Model (8) Using LLM Emotion (``Fear'' Class)}}}
\label{appendix:significance8}
\begin{tabular}{lccc} 
\toprule
\makecell[l]{F-Statistic\\(P-Value)} & \makecell{Disclosed\\\&\\Persistent} & \makecell{Disclosed\\\&\\Disappeared} & \makecell{Dark Money\\\&\\Persistent} \\[1mm]
\toprule
\multirow{2}{2.2cm}{Dark Money\\\& Disappeared} & 6.11** & 4.08** & 2.75*\\
& (0.013) & (0.043) & (0.097) \\[2mm]
\multirow{2}{2.2cm}{Dark Money\\\& Persistent} & 16.24*** & 1.77 \\
& ($5.6\mathrm{e}{-5}$) & (0.184)\\[2mm]
\multirow{2}{2.2cm}{Disclosed\\\& Disappeared} &  3.26*  \\ 
& (0.071) \\
\bottomrule\\
\multicolumn{4}{l}{F-Test between coefficients in Table~\ref{tab:reg_coeff} - Model (8), with }\\
\multicolumn{4}{l}{p-values reported in parentheses below each F-statistic.}\\
\multicolumn{4}{l}{$*$ $p <$ 0.1  $*$$*$ $p <$ 0.05  $*$$*$$*$ $p <$ 0.01} \\
\end{tabular} 
\end{table}
\begin{table}[ht]
\small
\centering
\caption{Estimated Differences in Sentiment by Accountability\\ \textbf{With Duplicate Ad Records Removed}}
\label{appendix:duplicates}
\begin{tabular}{lccc} 
\multicolumn{4}{c}{(1) All ads (2) All ads with fixed effects for advertiser type} \\[0.5mm]
\multicolumn{4}{c}{(3) Only ads from 501(c) organizations and corporations} \\[1mm] 

\toprule
& \makecell{(1)} & \makecell{(2)} & \makecell{(3)} \\[1mm]
\toprule
\multirow{2}{2.2cm}{Dark Money\\\& Disappeared} & $-$0.091** & $-$0.236*** & $-$0.239***\\
& (0.044) & (0.064) & (0.070) \\[2mm]
\multirow{2}{2.2cm}{Dark Money\\\& Persistent} & $-$0.049 & $-$0.164*** & $-$0.189*** \\ 
& (0.043) & (0.060) & (0.069) \\[2mm]
\multirow{2}{2.2cm}{Disclosed\\\& Disappeared} & 0.177** & 0.094 & 0.118 \\ 
& (0.076) & (0.071) & (0.092) \\[1mm]
\midrule
\multirow{2}{2.2cm}{\makecell[l]{Intercept\\(Disclosed \& Persistent)}} &  0.305*** & 0.369*** & 0.453*** \\ 
& (0.017)  & (0.013) & (0.056) \\[1mm]
\midrule
Number of Ads &  43,917 & 43,917 & 4,552 \\[2mm]

Adv Type Fixed Effects &  No & Yes & N/A \\

\bottomrule\\
\multicolumn{4}{l}{Replicates Table 1, Models (1)-(3). } \\[1mm]
\multicolumn{4}{l}{$*$ $p <$ 0.1  $*$$*$ $p <$ 0.05  $*$$*$$*$ $p <$ 0.01; T-Test for coefficient $\neq$ 0} \\[1mm]
\multicolumn{4}{l}{Intercept for (2) includes candidates as the omitted fixed effect} \\

\end{tabular} 
\end{table}
\begin{table}[ht]
\small
\centering
\caption{Estimated Differences in Sentiment by Accountability\\  \textbf{With Partial Disclosure Coded As Dark Money}}
\label{appendix:gray_money}
\begin{tabular}{lccc} 
\multicolumn{4}{c}{(1) All ads (2) All ads with fixed effects for advertiser type} \\[0.5mm]
\multicolumn{4}{c}{(3) Only ads from 501(c) organizations and corporations} \\[1mm] 

\toprule
& \makecell{(1)} & \makecell{(2)} & \makecell{(3)} \\[0.5mm]
\toprule
\multirow{2}{2.2cm}{Dark Money\\\& Disappeared} & $-$0.284*** & $-$0.126 & $-$0.411*** \\
& (0.100) & (0.134) & (0.099) \\[2mm]
\multirow{2}{2.2cm}{Dark Money\\\& Persistent} & $-$0.232*** & $-$0.106** & $-$0.338***\\ 
& (0.090) & (0.054) & (0.129)\\[2mm]
\multirow{2}{2.2cm}{Disclosed\\\& Disappeared} & 0.196 & 0.345*** & 0.089 \\ 
& (0.133) & (0.128) & (0.110) \\[1mm]
\midrule
\multirow{2}{2.2cm}{\makecell[l]{Intercept\\(Disclosed \& Persistent)}}  & 0.321*** & 0.456*** & 0.447*** \\ 
& (0.062) & (0.077) & (0.059) \\[1mm]
\midrule
Number of Ads & 430,044 & 430,044 & 46,959 \\[2mm]

Adv Type Fixed Effects &  No & Yes & N/A \\

\bottomrule\\
\multicolumn{4}{l}{Replicates Table 1, Models (1)-(3). } \\[1mm]
\multicolumn{4}{c}{$*$ $p <$ 0.1  $*$$*$ $p <$ 0.05  $*$$*$$*$ $p <$ 0.01; T-Test for coefficient $\neq$ 0} \\[1mm]
\multicolumn{4}{l}{Intercept for (2) includes candidates as the omitted fixed effect} \\

\end{tabular} 
\end{table}
\begin{table}[ht]
\small
\centering
\caption{Estimated Differences in Sentiment by Accountability\\  \textbf{With Ads From Outside Groups Only}}
\label{appendix:outside_groups}
\begin{tabular}{lccc} 
\multicolumn{4}{c}{(1) All ads (2) All ads with fixed effects for advertiser type} \\[0.5mm]
\multicolumn{4}{c}{(3) Only ads from 501(c) organizations and corporations} \\[1mm] 

\toprule
& \makecell{(1)} & \makecell{(2)} & \makecell{(3)} \\[1mm]
\toprule
\multirow{2}{2.2cm}{Dark Money\\\& Disappeared} & $-$0.093 & $-$0.265*** & $-$0.370*** \\
& (0.078) & (0.099)  & (0.087) \\[2mm]
\multirow{2}{2.2cm}{Dark Money\\\& Persistent} & $-$0.014 & $-$0.221** & $-$0.288** \\ 
& (0.114) & (0.100) & (0.126) \\[2mm]
\multirow{2}{2.2cm}{Disclosed\\\& Disappeared} & 0.380*** & 0.235** & 0.109 \\ 
& (0.081) & (0.105) & (0.091) \\[1mm]
\midrule
\multirow{2}{2.2cm}{\makecell[l]{Intercept\\(Disclosed \& Persistent)}}  & 0.112*** & 0.278*** & 0.390*** \\ 
& (0.024) & (0.097) & (0.045) \\[1mm]
\midrule
Number of Ads &  129,935 & 129,935 & 46,959 \\[2mm]

Adv Type Fixed Effects &  No & Yes & N/A \\

\bottomrule\\
\multicolumn{4}{l}{Replicates Table 1, Models (1)-(3). } \\[1mm]
\multicolumn{4}{c}{$*$ $p <$ 0.1  $*$$*$ $p <$ 0.05  $*$$*$$*$ $p <$ 0.01; T-Test for coefficient $\neq$ 0} \\[1mm]
\multicolumn{4}{l}{Intercept for (2) includes corporations as the omitted fixed effect} \\

\end{tabular} 
\end{table}

\clearpage

\begin{table}[ht]
\small
\centering
\caption{Estimated Differences in LLM Emotion Scores by Accountability}
\label{appendix:emotion}
\begin{tabular}{lcccccccc} 

\toprule
& \makecell{Anger} & \makecell{Disgust} & \makecell{Fear} & \makecell{Joy} & \makecell{Sadness} & \makecell{Surprise} & \makecell{Neutral}\\[1mm]
\toprule
\multirow{2}{3.5cm}{Dark Money\\\& Disappeared} & $-$0.046 & $-$0.036** &0.495** & $-$0.037* & $-$0.011**  & $-$0.025** & $-$0.340** \\
& (0.033) & (0.016) & (0.200) & (0.022) & (0.005) & (0.013) & (0.145)\\[2mm]
\multirow{2}{3.5cm}{Dark Money\\\& Persistent} & 0.050 & $-$0.017 & 0.158*** & $-$0.025 & 0.083 & $-$0.027*** & $-$0.222**\\ 
& (0.039) & (0.017) & (0.039) & (0.018) & (0.064) & (0.004) & (0.092) \\[2mm]
\multirow{2}{3.5cm}{Disclosed\\\& Disappeared} & $-$0.050** & $-$0.029* &0.082* & 0.032 & $-$0.005 & $-$0.006 & $-$0.023 \\ 
& (0.020) & (0.015) & (0.045) & (0.035) & (0.006) & (0.007) & (0.085) \\[1mm]
\midrule
\multirow{2}{3.5cm}{Intercept \\(Disclosed \& Persistent)} &  0.118*** & 0.051*** & 0.079*** & 0.062*** & 0.023*** & 0.045*** & 0.620*** \\ 
& (0.016)  & (0.014) & (0.012) & (0.017) & (0.004) & (0.003) & (0.046) \\[1mm]
\midrule
Number of Ads &  430,044 & 430,044 & 430,044 & 430,044 & 430,044 & 430,044 & 430,044 \\[2mm]

Advertiser Type Fixed Effects &  No & No & No & No & No & No & No \\

\bottomrule\\
\multicolumn{8}{l}{Regression coefficients represent estimated differences in LLM class probabilities for each emotion for each level of}\\
\multicolumn{8}{l}{advertiser accountability. Fear, shown here in column (3) is also in Table 1 - Model (8) in the main text.}\\
\multicolumn{8}{l}{$*$ $p <$ 0.1  $*$$*$ $p <$ 0.05  $*$$*$$*$ $p <$ 0.01; T-Test for coefficient $\neq$ 0} \\[1mm]

\end{tabular} 
\end{table}
\begin{table}[h!]
\small
\caption{Latent Dirichlet Allocation (LDA) Topic Modeling\\for Negative Ads (Lower Quartile of VADER Sentiment)}
\centering
\begin{tabular}{C{2cm}L{14cm}}

\toprule
Topic & \multicolumn{1}{c}{Words} \\
\midrule
\addlinespace

Fundraising & 0.029*``donat'' + 0.027*``democrat'' + 0.026*``voter'' + 0.024*``republican'' + 0.024*``campaign'' + 0.022*``elect'' + 0.022*``trump'' + 0.021*``help'' + 0.020*``attack'' + 0.017*``ad'' \\
\addlinespace
\midrule
\addlinespace
Healthcare & 0.033*``fight'' + 0.022*``healthcar'' + 0.021*``will'' + 0.021*``congress'' + 0.020*``work'' + 0.019*``big'' + 0.018*``peopl'' + 0.016*``drug'' + 0.016*``compani'' + 0.016*``veteran'' \\
\addlinespace
\midrule
\addlinespace
Voter Turnout & 0.265*``vote'' + 0.122*``nov'' + 0.118*``will'' + 0.112*``run'' + 0.098*``congress'' + 0.083*``district'' + 0.034*``state'' + 0.027*``repres'' + 0.022*``senat'' + 0.011*``david'' \\
\addlinespace
\midrule
\addlinespace
Tax Cuts & 0.064*``famili'' + 0.062*``work'' + 0.059*``tax'' + 0.035*``cut'' + 0.031*``fight'' + 0.024*``will'' + 0.021*``hard'' + 0.020*``michigan'' + 0.019*``pay'' + 0.019*``rais'' \\
\addlinespace
\midrule
\addlinespace
Brett Kavanaugh & 0.069*``senat'' + 0.052*``kavanaugh'' + 0.025*``women'' + 0.024*``brett'' + 0.024*``sexual'' + 0.023*``washington'' + 0.022*``vote'' + 0.021*``court'' + 0.018*``mike'' + 0.017*``assault'' \\
\addlinespace
\midrule
\addlinespace
Voter Turnout & 0.060*``poll'' + 0.057*``elect'' + 0.053*``vote'' + 0.052*``day'' + 0.046*``trump'' + 0.040*``find'' + 0.033*``place'' + 0.028*``sign'' + 0.025*``ballot'' + 0.020*``today'' \\
\addlinespace
\midrule
\addlinespace
Election Fraud & 0.090*``peopl'' + 0.058*``candid'' + 0.046*``ballot'' + 0.044*``bad'' + 0.043*``risk'' + 0.040*``absente'' + 0.040*``call'' + 0.035*``mail'' + 0.034*``request'' + 0.032*``travel'' \\
\addlinespace
\midrule
\addlinespace
N/A & 0.080*``vote'' + 0.073*``tax'' + 0.072*``governor'' + 0.045*``state'' + 0.040*``liber'' + 0.034*``novemb'' + 0.029*``cut'' + 0.025*``joe'' + 0.022*``job'' + 0.020*``immigr'' \\
\addlinespace
\midrule
\addlinespace
Guns \& Climate & 0.038*``gun'' + 0.037*``fight'' + 0.036*``trump'' + 0.027*``will'' + 0.023*``support'' + 0.022*``chang'' + 0.022*``stand'' + 0.022*``protect'' + 0.022*``communiti'' + 0.020*``climat'' \\
\addlinespace
\midrule
\addlinespace
Voter Turnout & 0.229*``vote'' + 0.069*``novemb'' + 0.027*``time'' + 0.027*``tuesday'' + 0.026*``earli'' + 0.021*``plan'' + 0.018*``voic'' + 0.016*``condit'' + 0.016*``elect'' + 0.016*``poll'' \\

\addlinespace

\bottomrule

\end{tabular}
\label{appendix:topics_negative}
\end{table}

\begin{table}
\small
\caption{Latent Dirichlet Allocation (LDA) Topic Modeling\\for Positive Ads (Upper Quartile of VADER Sentiment)}
\centering
\begin{tabular}{C{2cm}L{14cm}}

\toprule
Topic & \multicolumn{1}{c}{Words} \\
\midrule
\addlinespace

Brett Kavanaugh & 0.047*``senat'' + 0.044*``court'' + 0.039*``suprem'' + 0.037*``kavanaugh'' + 0.032*``tax'' + 0.027*``vote'' + 0.025*``will'' + 0.021*``brett'' + 0.016*``trump'' + 0.015*``judg'' \\
\addlinespace
\midrule
\addlinespace
Campaign Support & 0.032*``join'' + 0.022*``communiti'' + 0.019*``will'' + 0.017*``great'' + 0.016*``work'' + 0.015*``help'' + 0.015*``job'' + 0.013*``patagonia'' + 0.013*``grante'' + 0.012*``local'' \\
\addlinespace
\midrule
\addlinespace
Campaign Support & 0.023*``district'' + 0.023*``support'' + 0.022*``work'' + 0.022*``vote'' + 0.018*``peopl'' + 0.016*``congress'' + 0.015*``state'' + 0.014*``repres'' + 0.014*``republican'' + 0.014*``will'' \\
\addlinespace
\midrule
\addlinespace
Education & 0.048*``support'' + 0.036*``educ'' + 0.035*``school'' + 0.033*``public'' + 0.028*``vote'' + 0.027*``will'' + 0.026*``help'' + 0.020*``gift'' + 0.019*``fund'' + 0.018*``candid'' \\
\addlinespace
\midrule
\addlinespace
Voter Turnout & 0.057*``vote'' + 0.049*``voter'' + 0.048*``tuesday'' + 0.046*``poll'' + 0.041*``will'' + 0.033*``elect'' + 0.033*``counti'' + 0.029*``locat'' + 0.028*``ballot'' + 0.028*``place'' \\
\addlinespace
\midrule
\addlinespace
Healthcare & 0.086*``vote'' + 0.074*``care'' + 0.058*``health'' + 0.040*``novemb'' + 0.036*``congress'' + 0.030*``protect'' + 0.023*``will'' + 0.020*``afford'' + 0.019*``famili'' + 0.014*``peopl'' \\
\addlinespace
\midrule
\addlinespace
Voter Turnout & 0.045*``voter'' + 0.038*``vote'' + 0.036*``elect'' + 0.033*``will'' + 0.027*``poll'' + 0.026*``share'' + 0.025*``help'' + 0.024*``day'' + 0.020*``stand'' + 0.019*``locat'' \\
\addlinespace
\midrule
\addlinespace
Campaign Support & 0.049*``work'' + 0.038*``beto'' + 0.026*``will'' + 0.026*``join'' + 0.025*``famili'' + 0.025*``elect'' + 0.021*``win'' + 0.019*``meet'' + 0.018*``hard'' + 0.017*``campaign'' \\
\addlinespace
\midrule
\addlinespace
Fundraising & 0.030*``help'' + 0.022*``donat'' + 0.022*``will'' + 0.018*``support'' + 0.016*``veteran'' + 0.016*``secur'' + 0.014*``fight'' + 0.013*``win'' + 0.012*``race'' + 0.012*``democrat'' \\
\addlinespace
\midrule
\addlinespace
Voter Turnout & 0.042*``vote'' + 0.026*``support'' + 0.026*``presid'' + 0.023*``trump'' + 0.021*``today'' + 0.017*``day'' + 0.016*``will'' + 0.016*``campaign'' + 0.014*``chang'' + 0.013*``sign'' \\

\addlinespace

\bottomrule

\end{tabular}
\label{appendix:topics_positive}
\end{table}

\begin{table*}
\small
\caption{Examples of Ads and VADER Sentiment Scores}
\centering
\begin{tabular}{C{0.6cm}C{1.5cm}C{2cm}C{1.5cm}L{9.5cm}}

\toprule
Ad Id & Advertiser & Advertiser Type & Sentiment & \multicolumn{1}{c}{Ad Text} \\

\addlinespace
\midrule

\raisebox{-.1\normalbaselineskip}[0pt][0pt]{\rotatebox[origin=c]{90}{1474520922679905}} & Trump Make America Great Again Committee & Outside Group (PAC) & -0.782 & Our immigration system is in a state of COMPLETE CHAOS. Under current U.S. law, individuals can sponsor unlimited numbers of foreign immediate relatives for residency in the United States. THIS MUST END! Many of these chain migrants are not thoroughly vetted. This policy is a shameless Washington BETRAYAL of regular Americans whose safety is put at risk. You need to speak up so that the House and Senate KNOW what the American people think of this backward, anti-American policy. Please add your name to my OFFICIAL Presidential Petition to TERMINATE Chain Migration TODAY.
\\
\midrule
\addlinespace
\raisebox{-.05\normalbaselineskip}[0pt][0pt]{\rotatebox[origin=c]{90}{631573243903389}} & DCCC & Party & -0.741 & MUST-WATCH: Multiple women just accused Trump's Supreme Court nominee of sexual assault. Confirming Brett Kavanaugh to the Supreme Court would be horrifying for women -- period. That's why Democrats like Nancy Pelosi are fighting back. We CAN’T let Trump appoint an alleged serial abuser, overturn Roe v. Wade, and destroy women’s health. SIGN ON: PROTECT ROE V. WADE
\\

\addlinespace
\midrule
\addlinespace

\addlinespace
\raisebox{-.05\normalbaselineskip}[0pt][0pt]{\rotatebox[origin=c]{90}{555115378234868}} & Science for Humans and Freedom Institute & Outside Group (501(c) Org) & -0.542 & Google/YouTube regularly delete important videos, uploaded and linked by conservatives, creating MULTIPLE PERCEPTIONS of REALITY. The author and his regulars see a complete work, including videos. Liberals and others viewing the same piece later, see a black screen or dead link, and think that the author had no video evidence. \\
\addlinespace

\addlinespace
\midrule

\raisebox{-.05\normalbaselineskip}[0pt][0pt]{\rotatebox[origin=c]{90}{263752834246854}} & Need to Impeach & Outside Group (Super PAC) & -0.178 & It’s past time to impeach Donald Trump. He’s committed 9 impeachable offenses, including obstructing justice, violating the Emoluments Clause of the Constitution, conspiring to commit crimes against the U.S., endangering the peace and security of the U.S., and unconstitutionally imprisoning children. And now his personal lawyer just implicated President Trump in a criminal conspiracy to break campaign laws as a candidate. Demand impeachment now. \\

\midrule
\addlinespace

\addlinespace
\addlinespace
\addlinespace

\raisebox{-.01\normalbaselineskip}[0pt][0pt]{\rotatebox[origin=c]{90}{741934126159405}} & Republican National Committee & Party & 0.000 & President Trump and the GOP needs you to vote for the Republican ticket on November 6th! Find your election day polling location TODAY! \\
\addlinespace

\addlinespace
\addlinespace

\addlinespace
\midrule
\addlinespace

\raisebox{-.01\normalbaselineskip}[0pt][0pt]{\rotatebox[origin=c]{90}{355289031874657}} & Donald J. Trump For President, Inc. & Candidate & 0.151 & There are now less than 7 days left until the election. We are fighting the Hollywood liberal money machine, playing defense against the fake news, and battling UNPRECEDENTED obstruction within our own capital. With LESS than 7 days to go until the election, can I ask you, one of our most dedicated patriots, to contribute now to help us elect the allies I need to defend our agenda? \\

\addlinespace
\midrule

\raisebox{-.01\normalbaselineskip}[0pt][0pt]{\rotatebox[origin=c]{90}{157667358511395}} & Indivisible Project & Outside Group (501(c) Org) & 0.625 & It’s always hard to ask for money, so we’ll get right to the point: we’re hoping we can count on you to become a monthly donor to Indivisible. Our average donation is just under 30 dollars -- and the money we raise is what we use to power thousands of local groups knocking doors, making phone calls, and texting out the vote. Together we’re funding the tools and staff our local Indivisible groups need to succeed in 2018 and beyond: resisting the Trump agenda, electing progressive candidates up and down the ballot, and reshaping American democracy. If everyone reading this gave just a few bucks each month, we’d be able to sustain our work for years to come. Can you commit to giving monthly to Indivisible now? It just takes a minute to donate, but it makes a big difference for our movement. \\

\midrule
\addlinespace

\raisebox{-.01\normalbaselineskip}[0pt][0pt]{\rotatebox[origin=c]{90}{1037359523116149}} & Amy McGrath for Congress & Candidate & 0.981 & I decided to run for Congress because I want to fight for my children and my loved ones. I want to help shape a country in which the next generation receives the care they deserve, a country in which families know that their young ones are safe and protected. When I tell my kids about my run for Congress, I will recount this work as a time I am proud of. This ad, like all of the rest, is a positive one. I will continue to speak to my values and the issues I care about most. \\

\bottomrule

\end{tabular}
\label{appendix:ads}
\end{table*}

\end{document}